\documentclass[prl,floatfix,superscriptaddress,amsmath,amssymb,twocolumn]{revtex4}
\usepackage[latin1]{inputenc}
\usepackage{amsmath}
\usepackage{amssymb}
\usepackage{graphicx}
\usepackage{graphics}
\usepackage{color}
\usepackage{multirow}
\usepackage{subfigure}
\usepackage[export]{adjustbox}

\begin{document}

\title{Nematic and supernematic phases in Kagome quantum antiferromagnets under a magnetic field}

\author{Thibaut Picot}
\author{Didier Poilblanc}
\affiliation{Laboratoire de Physique Th\'eorique,  IRSAMC,
 CNRS and Universit\'e de Toulouse, UPS, F-31062 Toulouse, France}

\begin{abstract}
Optimizing {\it translationally invariant} infinite-Projected Entangled Pair States (iPEPS), we investigate  
the spin-2 Affleck-Kennedy-Lieb-Tasaki (AKLT) and spin-1 Heisenberg models on the Kagome lattice as a function of magnetic field.
We found that the magnetization curves offer a wide variety of compressible and incompressible phases.  
Incompressible nematic phases breaking the lattice $C_3$ rotation -- 
for which we propose simple qualitative pictures -- 
give rise to magnetization plateaux at reduced magnetization $m_z=5/6$ and $m_z=1/3$ for spin-2 and spin-1,  respectively, in addition to
the $m_z=0$ plateaux characteristic of zero-field gapped spin liquids. Moving away from the plateaux we observe a rich variety of compressible
superfluid nematic -- named ``supernematic" -- phases breaking spontaneously both point group and spin-U(1) symmetries, as well as a 
superfluid phase preserving lattice symmetries. 
We also identify the nature -- continuous or first-order -- of the various phase transitions.  
Possible connections to experimental spin-1 systems are discussed.

\end{abstract}
\maketitle


{\it Introduction} --
In the field of quantum magnetism, the magnetic field is a key control parameter which enables to explore exotic phase diagrams of spin systems.
In a spin ladder, one of the simplest spin system where spin-1/2 pair up into singlets on the rungs, the magnetic field, 
by playing the role of a chemical potential for triplet excitations on the rungs (triplons), can induce a transition to the spin-analog of the one-dimensional 
Luttinger Liquid 
(LL)~\cite{2leg_ladder}. Experimentally, synthetic gauge field (mimicking a magnetic field) in systems of bosonic atoms loaded on an optical lattice (formally equivalent to a
spin-1/2 quantum magnet) can be realized~\cite{cold_atoms}.
In two dimensions (2D), finite energy triplons can Bose-Einstein condense~\cite{BEC} and lead to a spin superfluid phase when the Zeeman coupling to the field overcomes the triplon energy. 
Field-induced
Bose-Einstein condensation (BEC) of triplons have indeed been experimentally observed in 2D quantum antiferromagnets (AFM) such as TlCuCl$_3$~\cite{TlCuCl},
Cs$_2$CuCl$_4$~\cite{CsCuCl} and BaCuSi$_2$O$_6$~\cite{BaCuSi}.
Another spectacular effect of the magnetic field, in the presence of magnetic 
frustration, is to give rise to magnetization plateaux in the magnetization curve. Such states with a fractional magnetization (when normalized w.r.t. the saturated value) are characterized by a gap towards spin excitations~: therefore, similarly to their charge analogs -- the Mott insulators -- they are incompressible. In celebrated examples such as the Shastry-Sutherland 
quantum spin-1/2 AFM
for strontium-copper-borate~\cite{StrontiumBorate}, or the spin-1/2 Kagome AFM (possibly) relevant for volborthite and vesignieite~\cite{volborthite}, it has been predicted that 
incompressible plateaux are stabilized under an applied
magnetic field by spontaneous breaking translation symmetry~\cite{StrontiumBorate,capponi}. 
In such cases, the stability of the plateau is 
generically associated to an {\it order by disorder} (OBD) mechanism~\cite{OBD} where the semi-classical long range (LR) spin order possesses a macroscopic 
entropy and maximizes quantum fluctuations. 
Whether plateau physics can occur in spin systems which preserve translation symmetry is still unknown
although such a translation invariant incompressible state can {\it formally} be constructed
based on an Affleck-Kennedy-Lieb-Tasaki (AKLT)~\cite{AKLT} framework involving resonating triplets 
polarized along the field~\cite{didier_AKLT_field}. In addition, search for plateaux
whose stabilization goes beyond the OBD mechanism is another motivation for our work.

\begin{figure}
\subfigure{\includegraphics[width=0.49\linewidth]{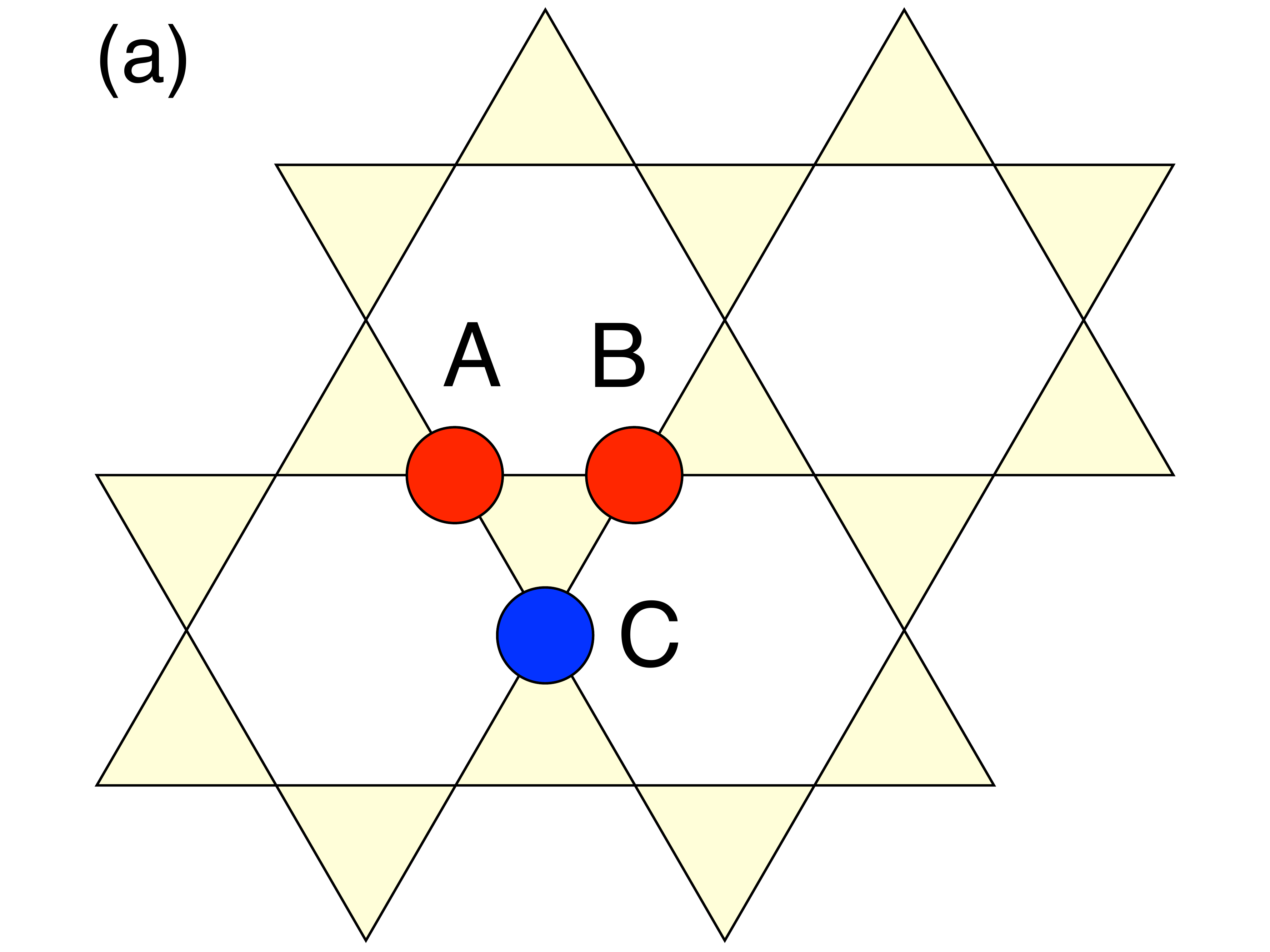}}
\subfigure{\includegraphics[width=0.49\linewidth]{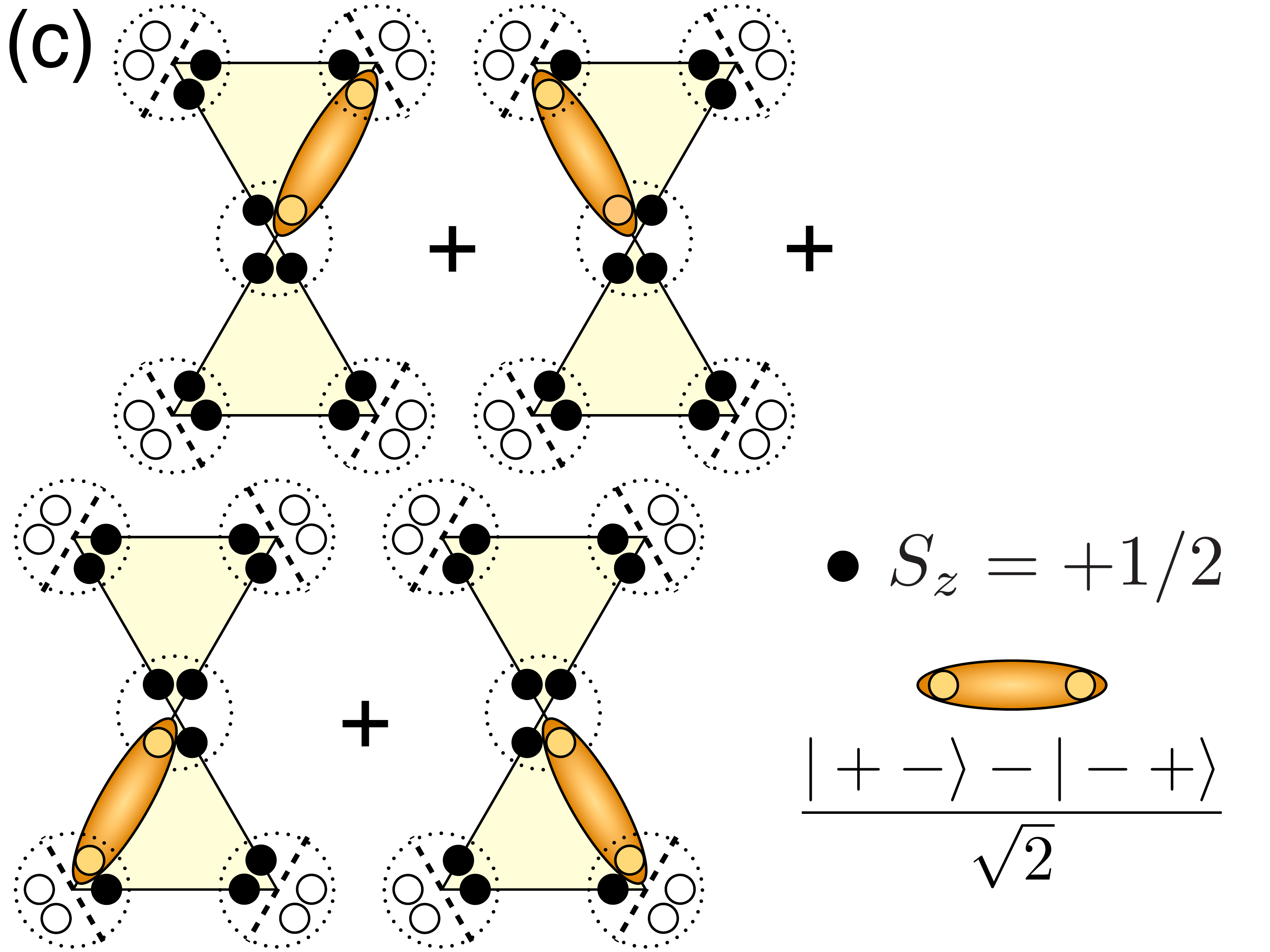}}
\subfigure{\includegraphics[width=0.32\linewidth]{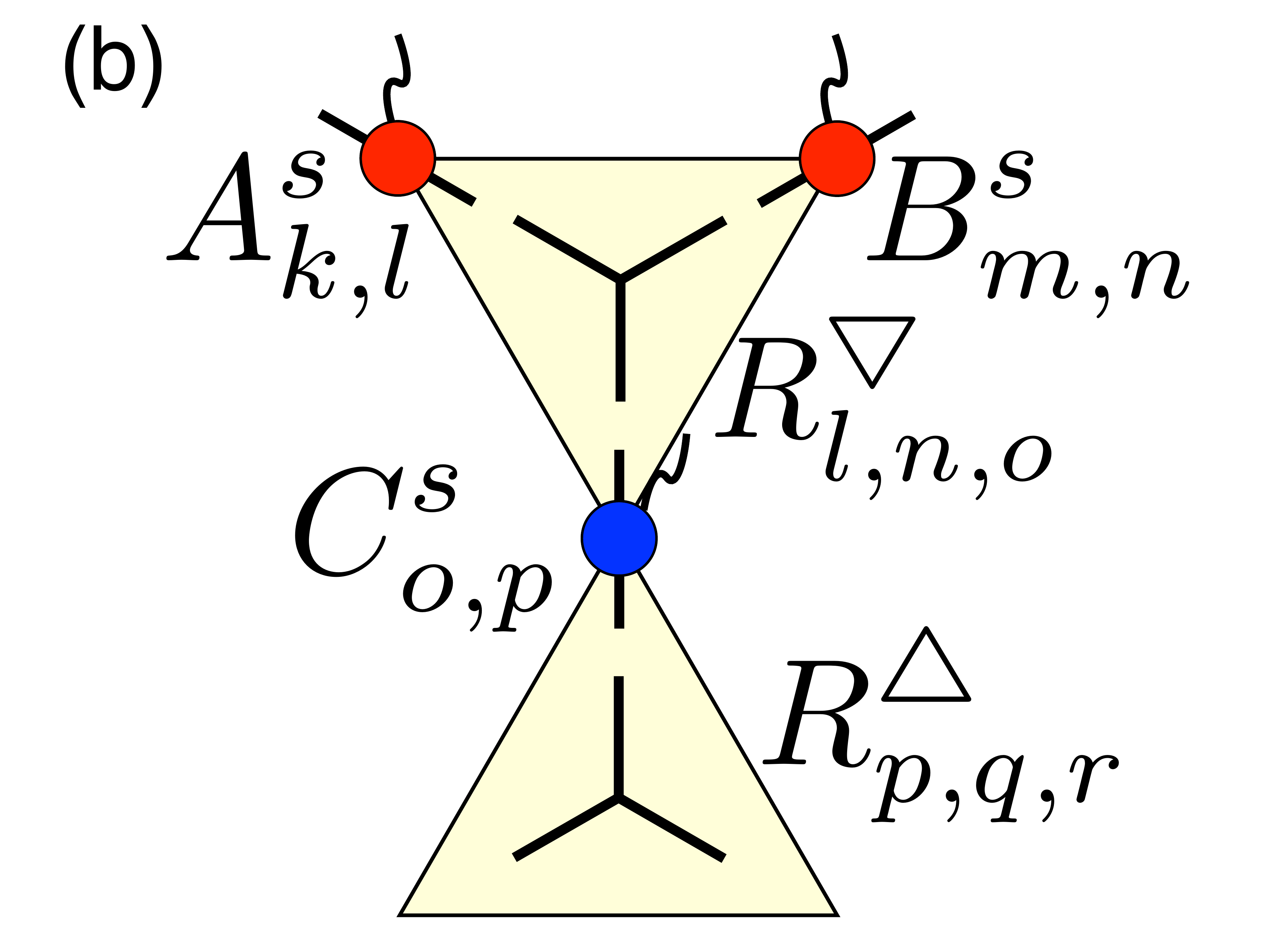}}
\subfigure{\includegraphics[width=0.32\linewidth]{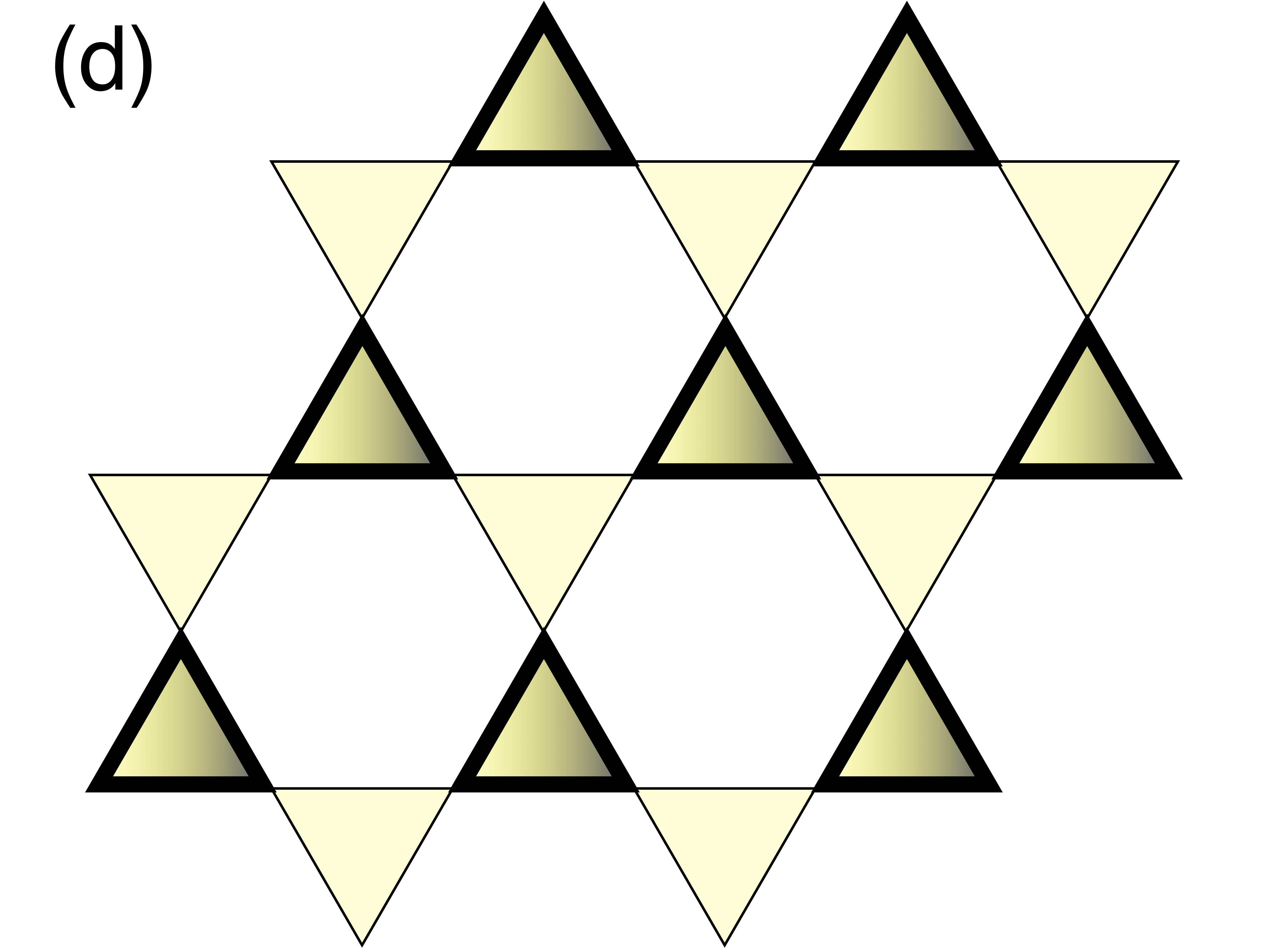}}
\subfigure{\includegraphics[width=0.32\linewidth]{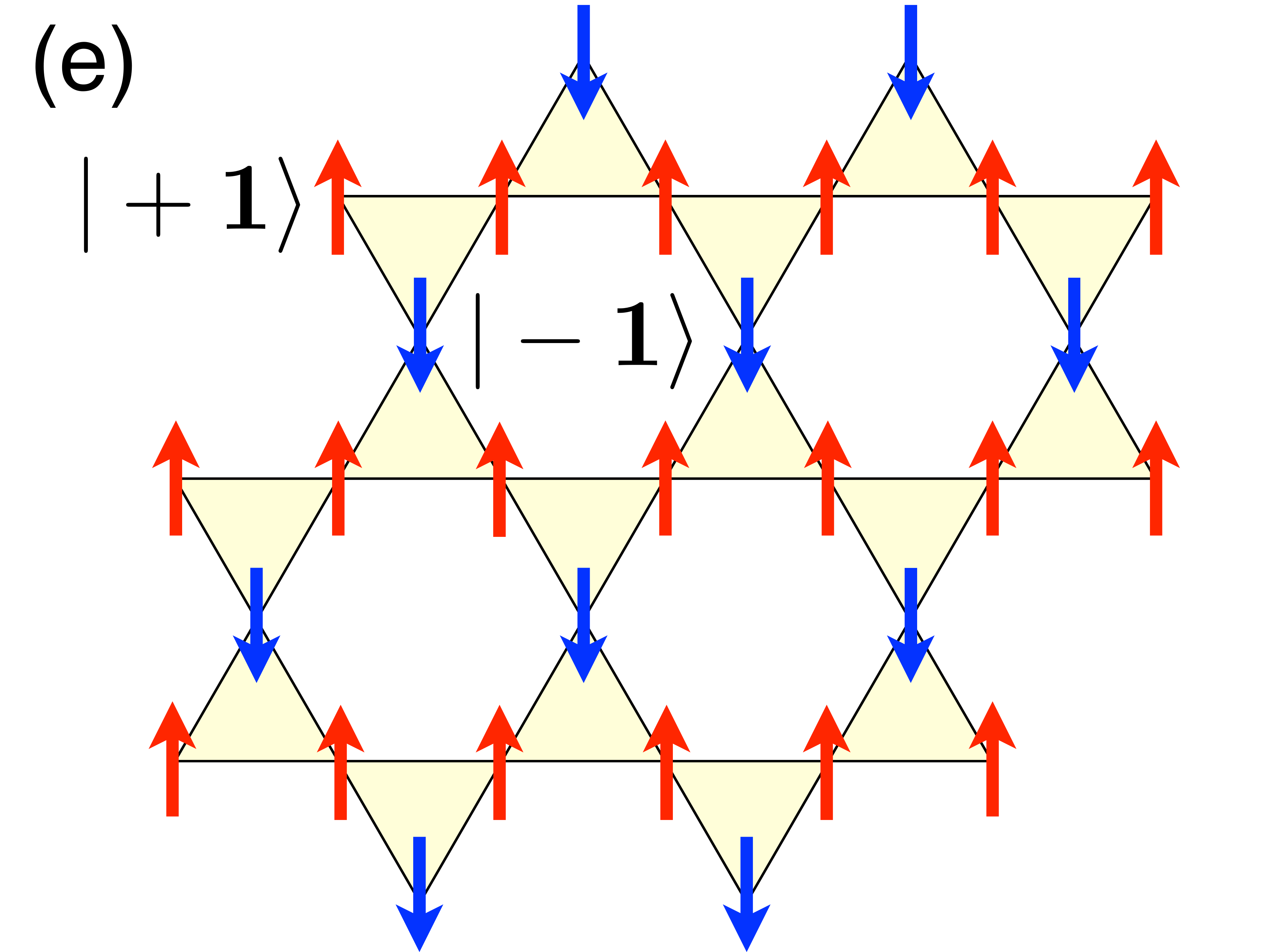}}
\caption{ (a) Kagome lattice and 3-site unit cell. (b) The five iPEPS tensors within the unit cell. (c) Schematic picture of the nematic phase 
in the AKLT model at $m_z=5/6$. 
Open (filled) circles are the ancilla spin-$1/2$ outside (inside) the unit cell and large dotted circles stand for the maps onto the spin-2 physical 
subspaces. (d) Simplex solid phase in the Heisenberg model at $h=0$. (e) Sketch of the semi-classical nematic phase in the spin-1 Heisenberg model at $m_z=1/3$.}
\label{fig:sketches}
\end{figure}

In this Letter we consider the spin-2 AKLT  
model
\begin{equation}
H_{\rm AKLT}=\sum_{\langle ij\rangle} P_{ij}^{(4)}-h\sum_i S_i^z
\end{equation}
 where $P_{ij}^{(4)}$ is the projector onto the $S=4$ subspace of  
the nearest-neighbor $\langle ij \rangle$ bond and the  
spin-1 quantum Heisenberg model 
\begin{equation}
H_H=\sum_{\langle ij \rangle}{\bf S}_i\cdot{\bf S}_j-h\sum_i S_i^z
\end{equation}
on the Kagome lattice in the presence of an applied magnetic field $h$ (chosen along the $z$ direction).
We use infinite-Projected Entangled Pair States (iPEPS)~\cite{iTEBD,MPS,Corner,CTMRG,TRG1,TRG2,TRG3} in the thermodynamic limit and find very rich phase diagrams. 
Although we restrict to {\it translationally invariant} states, we 
allow for non-equivalent A, B and C sites in the unit cell (see Fig.\ref{fig:sketches}(a)). Hence, point group symmetries 
may be spontaneously broken (while translation symmetry is preserved by construction). Such symmetry broken phases named ``nematics"
are characterized by fractional magnetization plateaux resulting from a gap towards spin 
excitations. Hence, the ``spin compressibility" (the derivative of the magnetization w.r.t. the field) vanishes and
these phases can be viewed as ``incompressible" (like ``solid" phases).
In addition, the Hamiltonian (spin) U(1) symmetry around the field direction may (independently) be spontaneously
broken resulting in superfluid phases -- or ``supernematic" 
phases whenever nematic order coexists with superfluidity.
In contrast, in the incompressible nematic phases the U(1) symmetry is preserved and the transverse magnetization vanishes 
on every site. 
A summary of the new phases found in this work and their symmetries -- compared to the more traditional solid and supersolid phases
not investigated here -- are shown in Table \ref{tab:symmetries}. 
 
Our iPEPS representation~\cite{didier_Kagome,PESS} involves three tensors $A^s_{k,l}$, $B^s_{m,n}$ and $C^s_{o,p}$ located on the lattice sites 
and two tensors $R^{\bigtriangledown}_{l,n,o}$ and $R^{\bigtriangleup}_{p,q,r}$ which connect the above three site-tensors on the down- and up-triangles, 
as shown in Fig.\ref{fig:sketches}(b). The five tensors are optimized through an imaginary time evolution procedure, starting from random initial states. Convergence w.r.t. the bond dimension $D$
was reached for, typically, $D=7$ (AKLT) and $D=15$ (spin-1 model). 
The various phases are then determined by analyzing (i) the magnetizations $\big<S^\mu_\alpha\big>$ on the 3 sites of the unit cell - both along ($\alpha=z$) 
and transverse ($\alpha=x,y$) to the magnetic field - and (ii) the bond-energy densities on the six non-equivalent 
bonds of the unit cell. For convenience, we define {\it reduced} magnetizations by normalizing the longitudinal 
and transverse components by the spin S, i.e. $m^\mu_z=\big<S_\mu^z\big>/S$ and 
$m^\mu_\perp=\sqrt{\big<S_\mu^x\big>^2+\big<S_\mu^y\big>^2}/S$ respectively.
The {\it total} (reduced) longitudinal magnetization is obtained by adding the (algebraic) contributions
from the 3 sites, $m_z=\sum_{\mu}\big<S_\mu^z\big>/3S$. Note that the total planar magnetization always vanishes, $\sum_{\mu}\big<S_\mu^\alpha\big>=0$
for $\alpha=x$ or $y$.
Spontaneous breaking of the $2\pi/3$-rotation (named $C_3$) is directly revealed by a difference in the longitudinal magnetization on some of the
 A, B or C sites and characterizes nematic (or supernematic) phases. 
 Note that we find that, generically, at least 2 of the 3 sites still have the same magnetization, i.e. a mirror symmetry is preserved.
Note also that the inversion (i.e. the $\pi$-rotation named $C_2$) transforming up- into down-triangles can be broken in the case of a ``simplex solid"
for which the bond energies differ on the two types of triangles.
 
 \begin{table}
\vspace{1em}
\begin{tabular}{|c|c|c|c|c|c|}
  \hline
  & Compressible & $U(1)$ & $C_3$ & $C_2$ & $\mathcal{G}_T$\\
  \hline
  Bose Glass & \checkmark & \checkmark & \checkmark & \checkmark & \checkmark \\
  \hline
  Valence Bond Solid (VBS) & $\times$ & \checkmark & \checkmark & \checkmark & \checkmark \\
  \hline
 Simplex Solid (SiSo) & $\times$ & \checkmark & \checkmark & $\times$ & \checkmark \\
  \hline
  Superfluid (SF) & \checkmark & $\times$ & \checkmark & \checkmark & \checkmark \\
  \hline
  Solid (S) & $\times$ & \checkmark & - & - & $\times$ \\
  \hline
  Nematic (N) & $\times$ & \checkmark & $\times$ & \checkmark & \checkmark \\
  \hline
  Supersolid (SS) & \checkmark & $\times$ & - & - & $\times$ \\
  \hline
  Supernematic (SN) & \checkmark & $\times$ & $\times$ & \checkmark & \checkmark \\
  \hline
\end{tabular}
\caption{Comparison between different field-induced phases in term of preserved (\checkmark) or broken ($\times$) 
point group ($C_3$ and $C_2$ rotations), translation group ($\mathcal{G}_T$) or spin-U(1) symmetries.
Compressibility  (\checkmark) or incompressibility ($\times$) is also indicated.}
\label{tab:symmetries}
\end{table}

\begin{figure}[t!]
\begin{center}
\subfigure{\includegraphics[trim=0mm 24mm 0mm 0mm, clip, width=\linewidth]{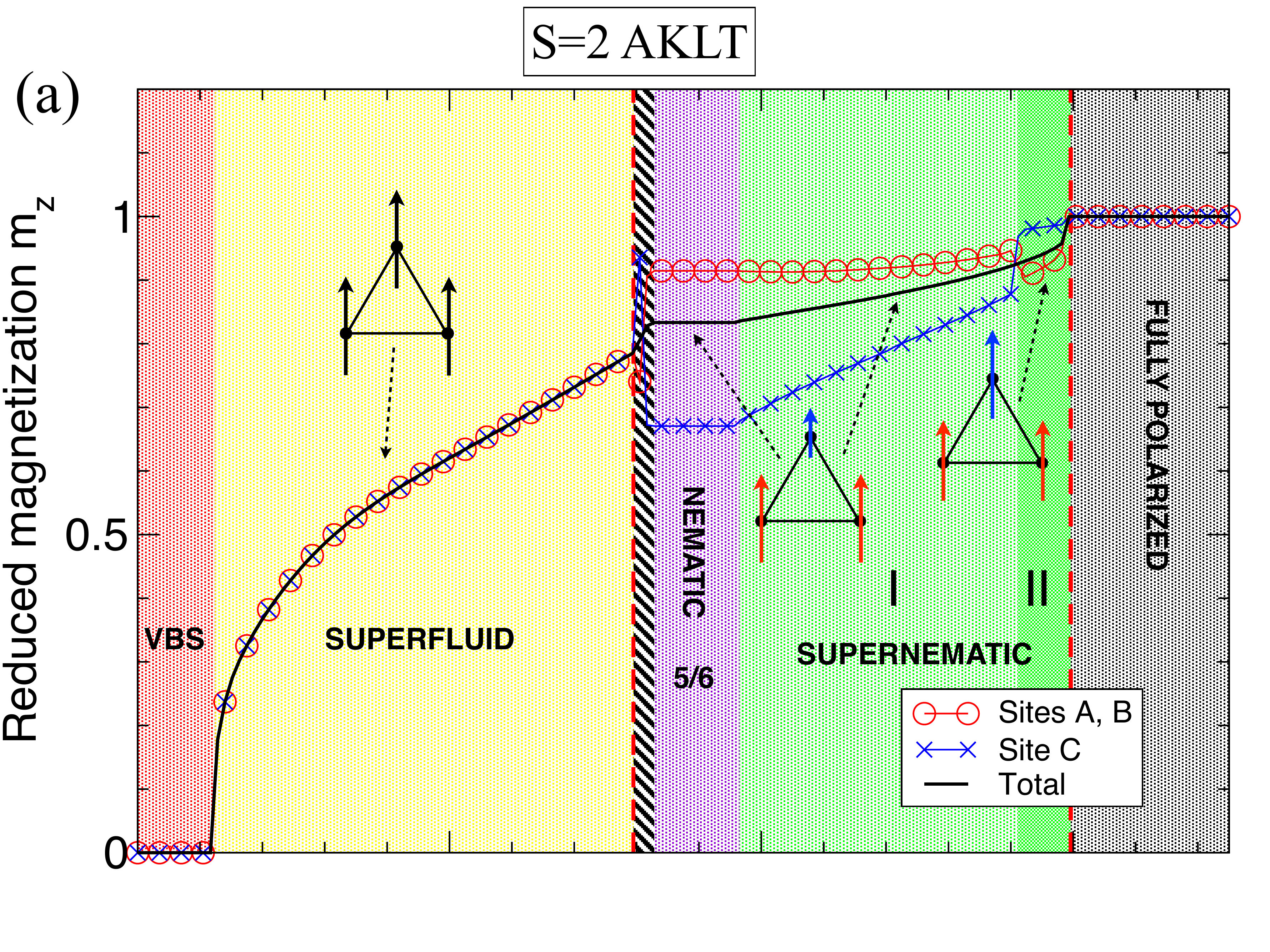}}
\subfigure{\includegraphics[trim=0mm 0mm 0mm 22mm, clip,width=\linewidth]{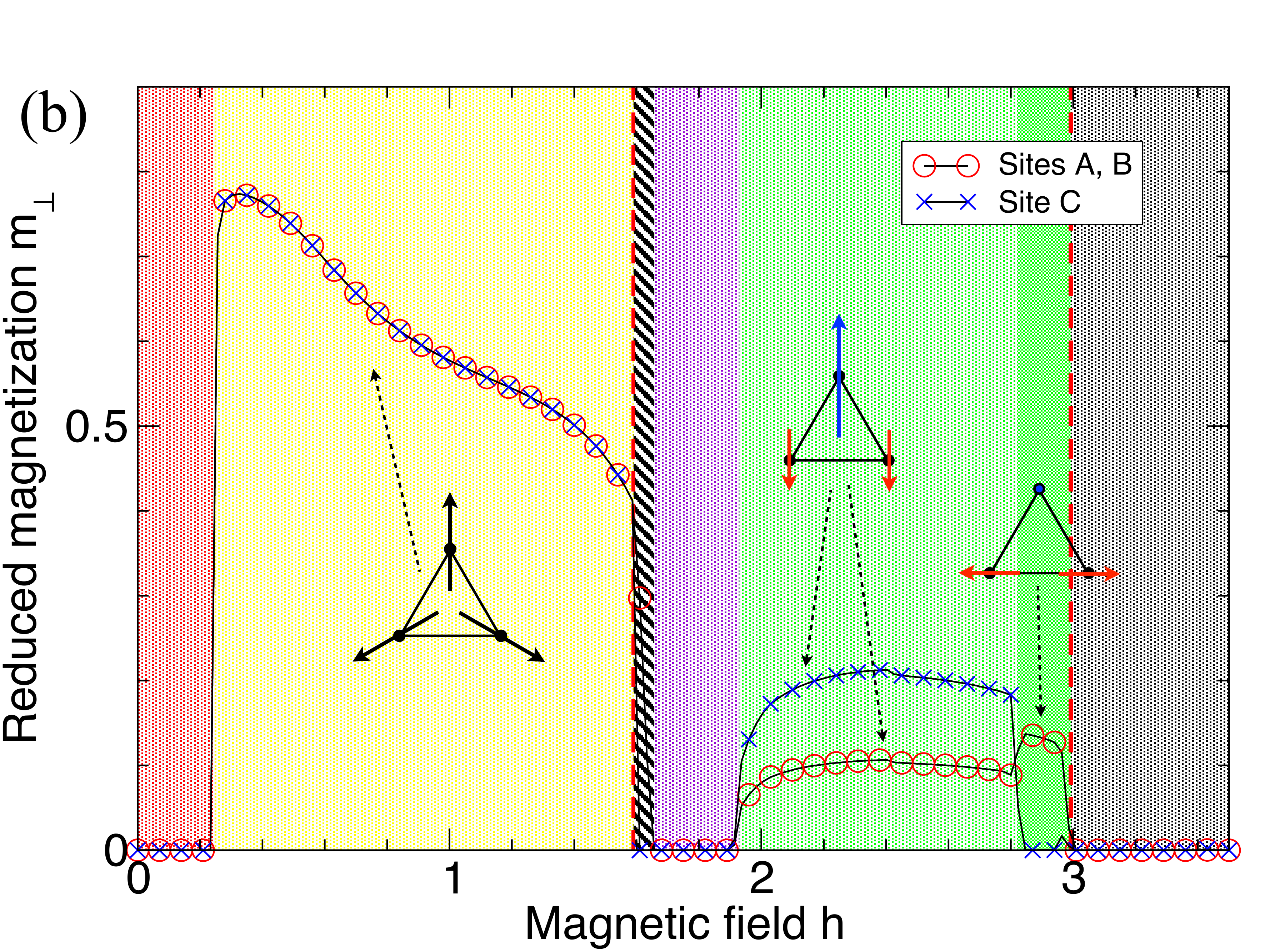}}
\end{center}
\caption{Phase diagram of the Kagome AKLT model vs magnetic field obtained from the analysis of the longitudinal and transverse magnetizations.  
First-order phase transitions are shown by red dashed lines. Reduced magnetization along the field (a) and transverse to
the field (b) are shown for the three sites of the unit cell. The total (reduced) magnetization 
is also shown in (a) as a continuous black line. Schematic pictures of the magnetization patterns (along and perpendicular to the field)
are provided for each phase. The $m_z=0$ and $m_z=5/6$ magnetization plateaux are characterized by $m_{\perp}=0$
in contrast to the superfluid and supernematic phases.}
\label{fig:aklt}
\end{figure}

{\it Spin-2 AKLT model} -- 
We first start with a description of the phase diagram of the S=2 AKLT model shown in Fig.~\ref{fig:aklt}.
Turning on the field, we first observe a $m_z=0$ magnetization plateau corresponding to the well-known {\it fully symmetric} singlet
Valence Bond Solid (VBS), stable up to a critical field of $h_c=\Delta_s/S\simeq 0.245$ where $\Delta_s$ is the spin gap of the AKLT 
(zero-field) phase~\cite{AKLT_gap}. The AKLT VBS is an exact $D=2$ PEPS~: on each site four spin-1/2 ancillas (or virtual states) are paired up into singlets 
with their neighbors and a map projects the virtual spin representation $\frac{1}{2}^{\otimes 4}$ onto the physical 
spin-2 subspace.  Above the critical field (where a more complex PEPS based upon the simplex tensors $R^{\bigtriangledown}$ and $R^{\bigtriangleup}$ is needed) we observe a compressible phase characterized by transverse components of the magnetization
at 120-degrees breaking  the spin-U(1) symmetry -- a characteristic of a superfluid (SF) phase -- and a uniform longitudinal magnetization.
At larger field we find a magnetization plateau at $m_z=5/6$ characteristic of an exotic incompressible phase.
This remarkable phase is a nematic which can be qualitatively pictured as a simple (D=3) PEPS as shown in Fig.~\ref{fig:sketches}(c)~: in the
unit cell all spin-1/2 ancillas are polarized ($S_z=+1/2$) except two which form a resonating singlet around site C. 
As in the usual $h=0$ AKLT state, local maps onto the spin-2 physical subspace are applied on every site. 
Such a simple picture enables to understand the broken point group symmetries of the nematic state~: the A and B sites remain equivalent,
as well as the 4 diagonal bonds and the 2 horizontal bonds.
The transition (or crossover) between the SF and the nematic phases is subtle and will be discussed later. 
Then, increasing the field further above a new critical value, transverse magnetic order smoothly appears while nematic order persists but 
with a finite compressibility $\partial m_z/\partial h$.  This supernematic I phase is characterized by equal collinear transverse magnetizations on the 
A and B sites. In contrast, in the supernematic II  phase at larger field the transverse magnetizations are opposite on A and B (and hence there is no transverse
magnetization on site C). Finally, we observe a jump of $m_z$ to the fully saturated value $m_z=1$ (1st order transition).

\begin{figure}[t!]
\begin{center}
\subfigure{\includegraphics[trim=0mm 24mm 0mm 0mm, clip, width=\linewidth]{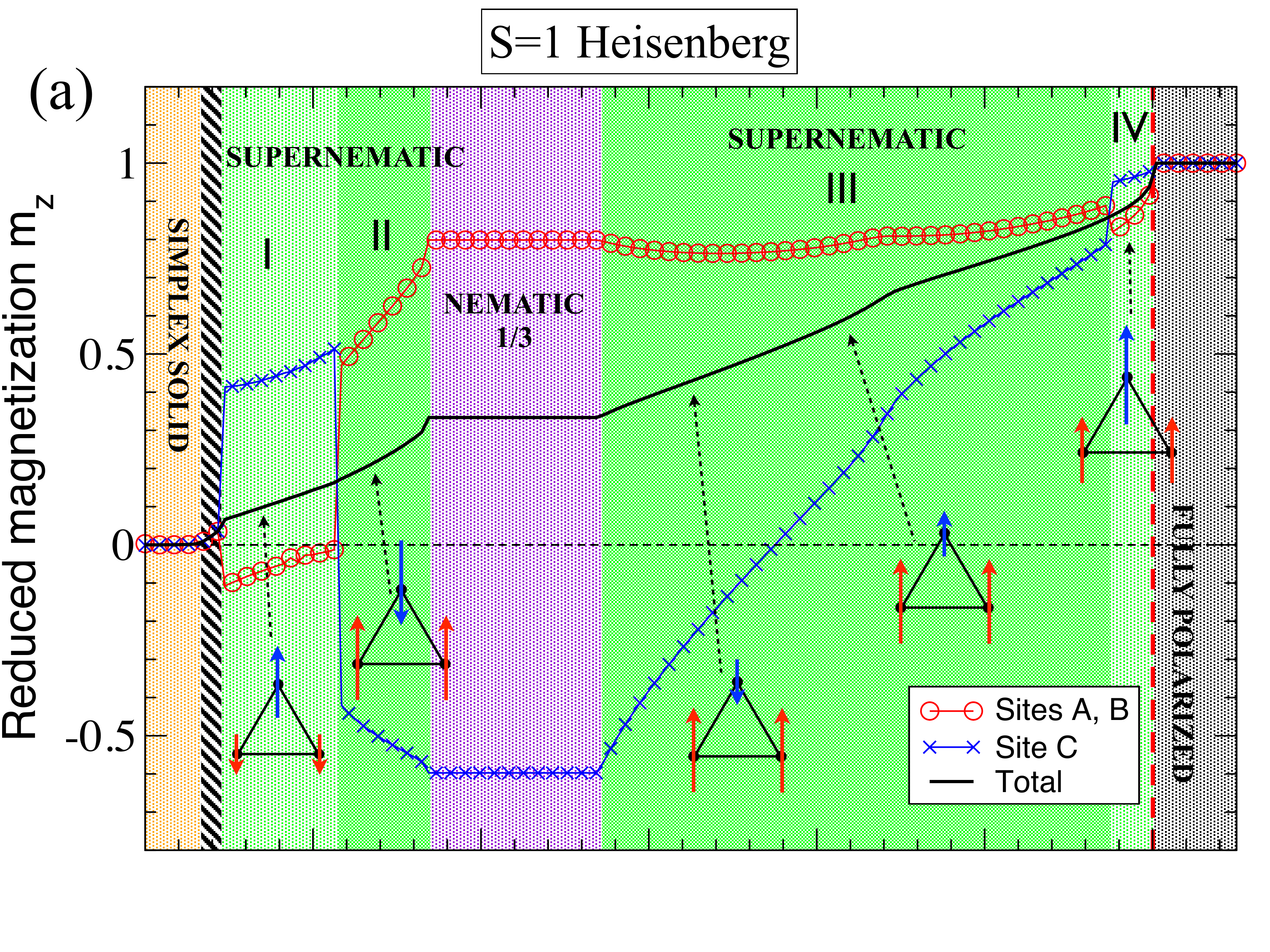}}
\subfigure{\includegraphics[trim=0mm 0mm 0mm 24mm, clip,width=\linewidth]{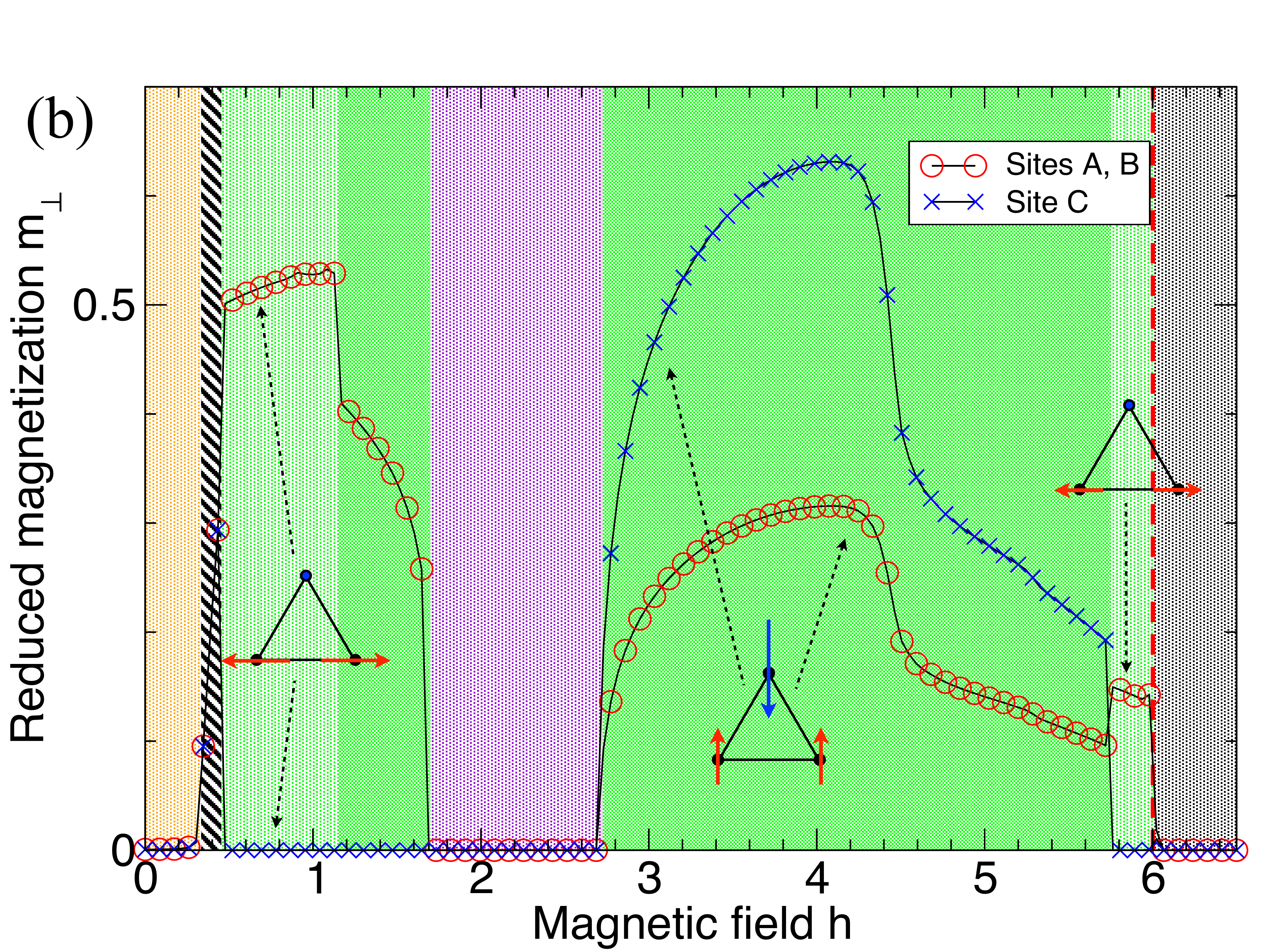}}
\end{center}
\caption{Same as Fig.~\protect\ref{fig:aklt} but for the Kagome spin-1 Heisenberg model. Magnetization plateaux
occur at $m_z=0$ and $m_z=1/3$, corresponding to simplex solid and nematic phases, respectively. Four types of compressible
supernematic phase (with superfluid order) appear in the neighborhood of the parent nematic phase. }
\label{fig:heis}
\end{figure}
\begin{figure*}[!t]
\includegraphics[trim=0mm 69mm 0mm 91mm, clip, width=\linewidth]{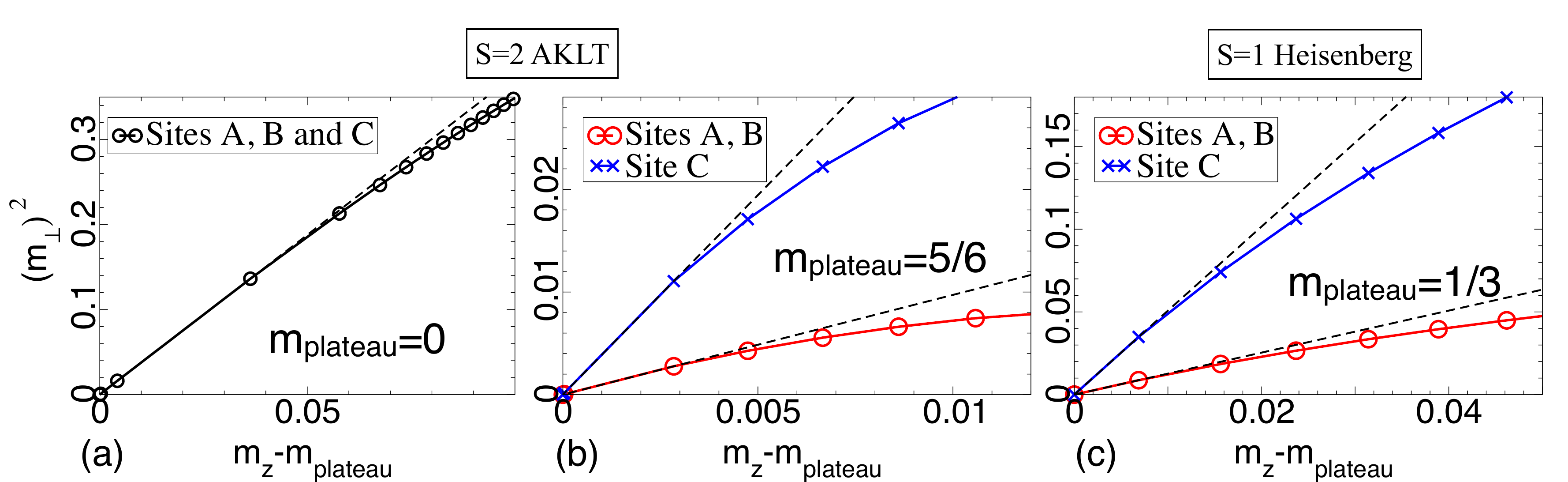}
\caption{Transverse magnetization squared $(m_\perp^\mu)^2$ on the A, B and C sites in the compressible phases 
in the vicinity of the magnetization plateaux (as indicated on plot) as a function of the deviation of the 
reduced total (longitudinal) magnetization
w.r.t. its fractional value on the plateau. Superfluid (a) and SN I (b) phases in the spin-2 AKLT model. (c)  SN III phase 
in the spin-1 AFM.
}
\label{fig:transition}
\end{figure*}

{\it Spin-1 Heisenberg model} --
We now turn to the description of the phase diagram of the spin-1 Heisenberg AFM as a function of the magnetic field, shown in Fig.~\ref{fig:heis}.
As for the AKLT model, we find a magnetization plateau at $m_z=0$, which is characteristic of a gapped phase. The corresponding 
(two-fold degenerate) simplex solid ground state (GS) depicted on Fig.~\ref{fig:sketches}(d)  is SU(2) invariant but breaks the ($C_2$) symmetry between up- and down-triangles due to different bond energies. 
Remarkably, the magnetization curve of the spin-1 AFM 
also shows a plateau at $m_z=1/3$. The corresponding incompressible state is a semi-classical ordered state, as pictured in Fig.~\ref{fig:sketches}(e),
with a favored direction (horizontal in the figure) of ferromagnetic chains ($m_z^A=m_z^B\simeq +1$) and N\'eel chains in the two other directions 
($m_z^C\simeq -1$). We expect this nematic state (breaking the $2\pi/3$-rotation) to be stabilized 
by quantum fluctuations via an OBD mechanism, in contrast to the AKLT plateau state. 
On each side of the magnetization plateau, we find supernematic phases with transverse components of the magnetization coexisting with the
longitudinal magnetic order inherited from the $m_z=1/3$ nematic phase. As for the spin-2 AKLT model, we find two types of supernematic phase 
depending whether the transverse components of the magnetization are opposite (SN~I, SN~II and SN~IV) or
collinear and equal (SN~III) on the A and B sites. Note that there is a transition between SN~I and SN~II (although they are of the same type) due to
a sudden change of sign of all the components of the magnetization along the field.

{\it Nature of phase transitions} --
Our phase diagrams offer a rich variety of phase transitions which can be classified according to the behavior
of the magnetization, a thermodynamic quantity, $m_z(h)=\partial E_{\rm GS} /\partial h$. 
(i) The transition from the VBS AKLT state to the SF is a continuous transition with $m_z\sim (h-h_c)^{1/2}$ and $m_\perp\sim m_z^{1/2}$ as shown in Fig.~\ref{fig:transition}(a). 
We can view it as a traditional BEC of triplons. (ii) We observe a second class of (2nd order) continuous transitions 
on the right edges of the two magnetization plateaux clearly characterized by $m_z-m_{\rm plateau}\propto (h-h_c)$
and $m^\mu_\perp\propto (m_z-m_{\rm plateau})^{1/2}$, as shown in Fig.~\ref{fig:transition}(b,c), where $m_{\rm plateau}$ corresponds to the fractional magnetization 5/6 and 1/3 of the plateaux of the AKLT and
Heisenberg models, respectively. 

(iii) The transitions on the left edges of the plateaux are more unconventional. For the spin-1 Heisenberg AFM, the transition between SN II  and 
the nematic phase may be weakly first order, or $m_z(h)$ may have a vertical slope. For the AKLT model there is a narrow intermediate region 
between the superfluid and the nematic phase which might correspond to another supernematic phase (similar to SN IV). We also identify a narrow interval
of superfluid phase between the simplex solid and SN I in the spin-1 AFM involving at least a first order transition. 
Also, we observe that the transition to the fully polarized state is generically of first order with a jump of the total magnetization (and of other observables).

(iv) Transitions between the supernematic phases, i.e. between SN I and SN II (in both phase diagrams) and between SN III and SN IV
are remarkably smooth. Indeed, they do not seem to involve a discontinuity of the {\it slope} of $m_z(h)$ (a second derivative of the energy)
-- strictly speaking they would correspond to 3rd order phase transitions -- although observables like the individual site magnetizations are
all discontinuous. 

{\it Discussions and conclusions} --
In this Letter, we study the very rich phase diagrams under a magnetic field of two Kagome quantum antiferromagnets  
described by the S=2 AKLT and the S=1 Heisenberg models. Besides known phases, such as superfluid, valence bond solid or fully polarized phases, we establish the existence of new remarkable field-induced phases, namely, {\it nematic} phases which break 
$2\pi/3$-rotation and preserve all other symmetries, and {\it supernematic} phases which break, in addition, the spin-U(1) symmetry. $Nematic$ phases are incompressible and, therefore, give rise to magnetization plateaux -- at $m_z=5/6$ and $m_z=1/3$ for S=2 and S=1, respectively. 
A wide variety of phase transitions are revealed and discussed. We also find that the zero-field GS of the S=1 AFM (using a {\it translationally-invariant} ansatz) is gapped and trimerized ($\pi/2$-rotation symmetry breaking), realizing  a new form
of incompressible state named {\it simplex solid}. 
Also, note that Tensor Network techniques, similar to the numerical method used in our work, have been used more recently in quantum chemistry~\cite{quantum_chemistry} and have potential applications in high energy physics like QED~\cite{QED}, or quantum gravity~\cite{quantum_gravity}.

It is interesting to discuss here the relevance of this work to experiments on some Nickelate or Vanadate compounds consisting of (weakly coupled) S=1 Kagome AFM layers. 
Reduced spin fluctuations (compared to spin-1/2 analogs) and possible existence of single-ion anisotropy might restrict the observation of a spin liquid behavior in such systems. 
For example, BaNi$_3$(OH)$_2$(VO$_4$)$_2$ shows a glassy behavior at low temperature whose origin is still unknown~\cite{exp1}.
Nevertheless, experimental behaviors suggestive of a spin gap have been seen in 
YCa$_3$(VO)$_3$(BO$_3$)$_4$~\cite{exp2} and KV$_3$Ge$_2$O$_9$~\cite{exp3} and Kagome compounds based on spin-1 V$^{3+}$ ions.
Although current single crystals might still contain a sizable amount of impurities~\cite{exp2} or competing  ferromagnetic interactions might be present~\cite{exp3}, 
such materials might nevertheless be good candidate to observe the exotic physical behaviors described here.

{\bf Acknowledgment} --  We acknowledge the
 NQPTP ANR-0406-01 grant (French Research Council) for support 
and the CALMIP Hyperion Cluster (Toulouse) for CPU-time. We
also thank Ignacio Cirac, Norbert Schuch and Sylvain Capponi for useful comments. 

Note added.-- Upon finalizing the manuscript we noticed two
recent preprints on the zero-field
spin-1 Kagome Heisenberg model reaching similar conclusions~\cite{lauchli,liu}.


\begin{thebibliography}{99}
\bibitem{2leg_ladder} C.A. Hayward, D. Poilblanc, and L. Levy, Phys. Rev. B {\bf 54}, R12649(R) (1996);
Ch. R\"uegg, K. Kiefer, B. Thielemann, D. F. McMorrow, V. Zapf, B. Normand, M. B. Zvonarev, P. Bouillot, C. Kollath, T. Giamarchi, S. Capponi, D. Poilblanc, D. Biner, K. W. Kr\"amer, Phys. Rev. Lett. {\bf 101}, 247202 (2008). 

\bibitem{cold_atoms} J. Simon, W. S. Bakr, R. Ma, M. E. Tai, P. M. Preiss and M. Greiner, Nature {\bf 472}, 307-312 (2011).

\bibitem{BEC} Thierry Giamarchi, Christian R\"uegg and Oleg Tchernyshyov, Nature Physics {\bf 4}, 198 - 204 (2008).
 
\bibitem{TlCuCl} A. Oosawa, M. Ishii and H. Tanaka, J. Phys. Condens. Matter 11, 265 (1999); T. Nikuni, M. Oshikawa, A. Oosawa, and H. Tanaka, Phys. Rev. Lett. {\bf 84}, 5868 (2000).

\bibitem{CsCuCl} T. Radu, H. Wilhelm, V. Yushankhai, D. Kovrizhin, R. Coldea, Z. Tylczynski, T. Luhmann, and F. Steglich, Phys. Rev. Lett. {\bf 95}, 127202 (2005)

\bibitem{BaCuSi} Ch. R\"uegg, D.F. McMorrow, B. Normand, H.M. Ronnow, S.E. Sebastian, I.R. Fisher, C.D. Batista, S.N. Gvasaliya, Ch. Niedermayer, and J. Stahn, Phys. Rev. Lett. {\bf 98}, 017202 (2007).

\bibitem{StrontiumBorate} 
H. Kageyama, K. Yoshimura, R. Stern, N. V. Mushnikov, K. Onizuka, M. Kato, K. Kosuge, C. P. Slichter, T. Goto, and Y. Ueda, Phys. Rev. Lett. {\bf 82}, 3168 (1999);
Y. H. Matsuda, N. Abe, S. Takeyama, H. Kageyama, P. Corboz, A. Honecker, S. R. Manmana, G. R. Foltin, K. P. Schmidt, F. Mila, Phys. Rev. Lett. {\bf 111}, 137204 (2013) and references therein.

\bibitem{volborthite} Yoshihiko Okamoto, Masashi Tokunaga, Hiroyuki Yoshida, Akira Matsuo, Koichi Kindo, 
and Zenji Hiroi, Phys. Rev. B 83, 180407(R) (2011).

\bibitem{capponi} 
Sylvain Capponi, Oleg Derzhko, Andreas Honecker, Andreas M. L\"auchli, and Johannes Richter, 
Phys. Rev. B 88, 144416 (2013); 
Satoshi Nishimoto, Naokazu Shibata, and Chisa Hotta,
Nature Communications  {\bf 4}, 2287 (2013) and references therein.


\bibitem{OBD} J. Villain, R. Bidaux, J.-P. Carton, and R. Conte, J. Phys. France {\bf 41}, 1263-1272 (1980).

\bibitem{AKLT} Ian Affleck, Tom Kennedy, Elliott H. Lieb, and Hal Tasaki, Phys. Rev. Lett. {\bf 59}, 799 (1987).

\bibitem{didier_AKLT_field} Didier Poilblanc, Norbert Schuch, and J. Ignacio Cirac,
Phys. Rev. B {\bf 88}, 144414 (2013).

\bibitem{MPS} J. Jordan, R. Orus, G. Vidal, F. Verstraete, J. I. Cirac, Phys. Rev. Lett. {\bf101}, 250602 (2008).

\bibitem{iTEBD} Roman Orus, and Guifre Vidal, Phys. Rev. B {\bf 78}, 155117 (2008).

\bibitem{Corner} T. Nishino, K. Okunishi, J. Phys. Soc. Jpn. {\bf 65} pp. 891-894 (1996); T. Nishino, K. Okunishi, J. Phys. Soc. Jp. {\bf 66}, 3040 (1997)

\bibitem{CTMRG} Roman Orus, and Guifre Vidal, Phys. Rev. B {\bf 80}, 094403 (2009).

\bibitem{TRG1} M. Levin, C. P. Nave, Phys. Rev. Lett. {\bf 99}, 120601 (2007)

\bibitem{TRG2} Z. Y. Xie, H. C. Jiang, Q. N. Chen, Z. Y. Weng, T. Xiang, Phys. Rev. Lett. {\bf 103}, 160601 (2009); H. H. Zhao, Z. Y. Xie, Q. N. Chen, Z. C. Wei, J. W. Cai, T. Xiang, Phys. Rev. B {\bf 81}, 174411 (2010)

\bibitem{TRG3} Z. Y. Xie, J. Chen, M. P. Qin, J. W. Zhu, L. P. Yang, T. Xiang, Phys. Rev. B {\bf 86}, 045139 (2012)

\bibitem{didier_Kagome} Norbert Schuch, Didier Poilblanc, J. Ignacio Cirac, and David Perez-Garcia,
Phys. Rev. B 86, 115108 (2012); Didier Poilblanc and Norbert Schuch, Phys. Rev. B {\bf 87}, 140407 (2013).

\bibitem{PESS} Z. Y. Xie, J. Chen, J. F. Yu, X. Kong, B. Normand, and T. Xiang, Phys. Rev. X {\bf 4}, 011025 (2014).

\bibitem{AKLT_gap} Artur Garcia-Saez, Valentin Murg, and Tzu-Chieh Wei, Phys. Rev. B {\bf 88}, 245118 (2013).

\bibitem{quantum_chemistry} G. K.-L. Chan, J. J. Dorando, D. Ghosh, J. Hachmann, E. Neuscamman, H. Wang, T. Yanai, Prog. Theor. Chem. and Phys., 18, 49 (2008).

\bibitem{QED} S. K\"uhn, J. I. Cirac and M.-C. Ba\~nuls Phys. Rev. A {\bf 90}, 042305.

\bibitem{quantum_gravity} M. Van Raamsdonk, arXiv:0907.2939; M. van Raamsdonk, Gen. Rel. Grav. {\bf 42} 2323- 2329 (2010) and Int. J. Mod. Phys. D {\bf 19} 2429-2435 (2010); J. Maldacena, L. Susskind, arXiv:1306.0533.


\bibitem{exp1} Danna E. Freedman, Robin Chisnell, Tyrel M. McQueen, Young S. Lee, Christophe Payen, and Daniel G. Nocera, Chem. Commun. {\bf 48}, 64-66, (2012).

\bibitem{exp2} 
Wojciech Miiller, Mogens Christensen,, Arfhan Khan, Neeraj Sharma, Ren\'e B. Macquart, Maxim Avdeev, Garry J. McIntyre, Ross O. Piltz, and Chris D. Ling, Chem. Mater. {\bf 23}, 1315 (2011).

\bibitem{exp3} Shigeo Hara, Hirohiko Sato, and Yasuo Narumi, J. Phys. Soc. Jpn. {\bf 81}, 073707 (2012).


\bibitem{lauchli} Hitesh J. Changlani and Andreas M. L\"auchli, arXiv:1406.4767.

\bibitem{liu} Tao Liu, Wei Li, Andreas Weichselbaum, Jan von Delft, and Gang Su, arXiv:1406.5905.
 
\end{thebibliography}
\end{document}


\title{Nematic and supernematic phases in kagome quantum antiferromagnets under a magnetic field: \textit{supplemental material}}

\author{Thibaut Picot}
\author{Didier Poilblanc}
\affiliation{Laboratoire de Physique Th\'eorique,  IRSAMC,
 CNRS and Universit\'e de Toulouse, UPS, F-31062 Toulouse, France}

\maketitle

\section{Simple Update}

\subsection{Description}

\begin{figure}
\includegraphics[trim=47mm 30mm 95mm 40mm, clip, scale=0.5]{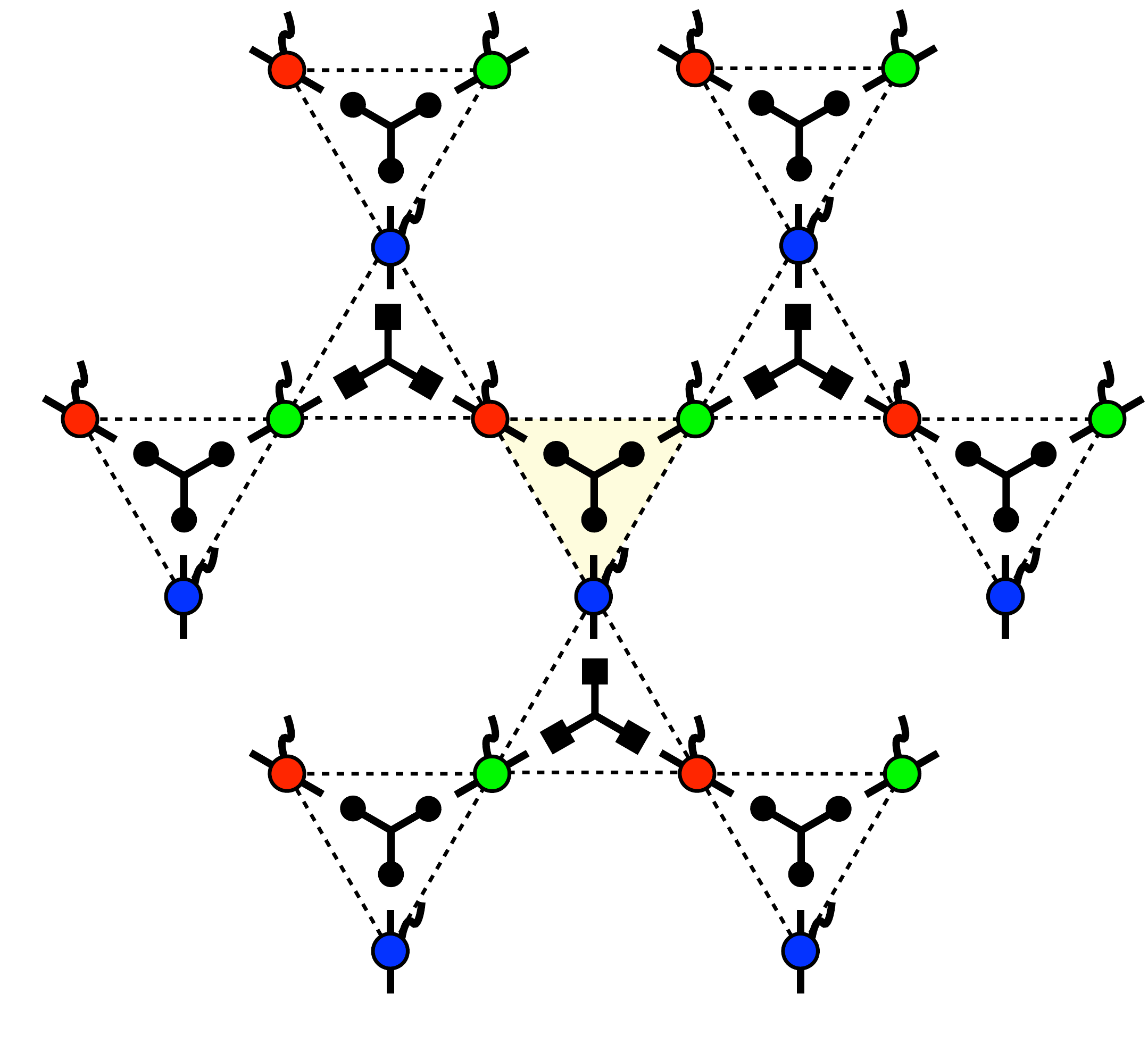}
\caption{Cluster used in the Simple Update approximation. The red, green and blue site tensors picture the A, B and C tensors. The simplex tensors with circle (square) ends depict the down (up) triangle simplex tensors.}
\label{fig:bethe}
\end{figure}

Tensor Networks in 2 dimensions in the thermodynamic limit, called \textit{infinite} Projected Entangled Pair State (iPEPS), can be optimized using 
Full Update or Simple Update schemes. The Full Update is more accurate but requires the calculation of the surrounding system of the unit cell -- called environment -- at every step. The (full) calculation of the environment can be performed using a Matrix Product State (MPS)-based approach~\cite{iTEBD,MPS}, a Corner Transfer Matrix Renormalization Group (CTMRG)~\cite{Corner,CTMRG} or a Coarse Graining Tensor Renormalization Group (CGTRG) method~\cite{TRG1,TRG2,TRG3} (or any related technic). In contrast, in the Simple Update, the tensor optimization used in this Letter, the (full) computation of the environment is not done during the optimization stage.
The resulting optimized tensor may have a lower accuracy, but the method allows to go to higher values of the bond dimension D.
Our update, using the imbedded cluster depicted in Fig.~\ref{fig:bethe}, is more reliable if quantum fluctuations do not extend much beyond the unit cell
(it becomes exact if one considers a Bethe lattice). 
In fact, the physics of the kagome antiferromagnet with integer spins ($S=1$ and $S=2$ in this Letter) involves physical processes mostly within the triangles instead of the hexagons, since an isolated triangle has a unique singlet ground state (instead of a doublet for S=1/2). Furthermore, our iPEPS representation considers a larger cluster (the focused triangle with its three neighboring triangles as shown in Fig.~1), compared to the square lattice where only two sites are probed.
Lastly, spin systems with spin larger than 1/2  generically exhibit less important quantum fluctuations. 

\subsection{Algorithm}

The optimization of the tensors in the Simple Update is based on $\lambda$ matrices which are located on the external bonds of the triangle. In one dimensional system, these $\lambda$ matrices contain the singular values of a decomposition of two neighboring sites, and connect these two sites. In two dimensions, they are an approximation of the outside system, whereas the $R^{\bigtriangledown(\bigtriangleup)}$ simplex tensor connect the three sites inside the focused triangle.

\begin{figure}
\includegraphics[trim=0mm 90mm 0mm 60mm, clip, scale=0.4]{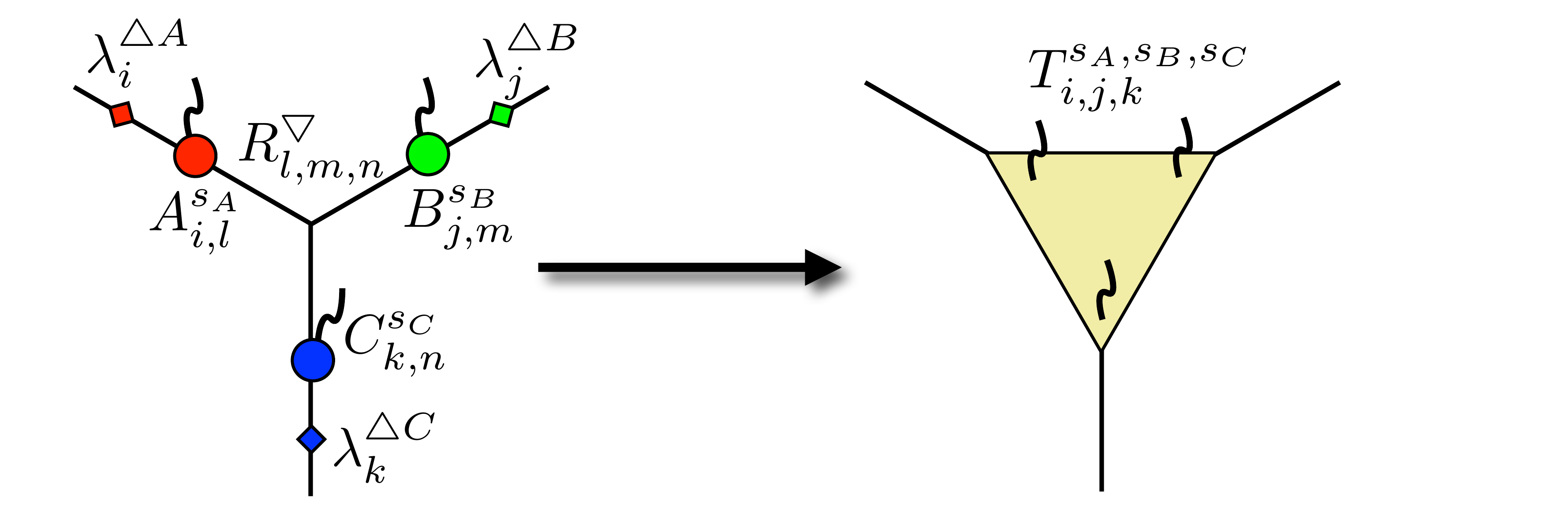}
\caption{$T$ tensor contraction, including $\lambda^{\bigtriangleup}$ diagonal matrices on the down triangle.}
\label{fig:algo}
\end{figure}

In the following, the algorithm for one optimization of the tensors on down triangles is described~\cite{PESS}.
First, we contract the $T$ tensor on the triangle, as shown in Fig.~\ref{fig:algo},
\[
T^{S_A,S_B,S_C}_{i,j,k}=\sum_{l,m,n}A^{S_A}_{i,l}B^{S_B}_{j,m}C^{S_C}_{k,n}R^{\bigtriangledown}_{l,m,n}\lambda^{\bigtriangleup A}_i\lambda^{\bigtriangleup B}_j\lambda^{\bigtriangleup C}_k
\]

Then, we apply the imaginary time evolution operator $U=\exp\left(-\delta \tau H^{\bigtriangledown}\right)$ on $T$: 
\[V^{\vec{S}}_{i,j,k}=\sum_{\vec{S}'}U_{\vec{S},\vec{S}'}T^{\vec{S}'}_{i,j,k}\]
where $\vec{S}$ is a vector, whose components are $S_A$, $S_B$ and $S_C$, and $\delta\tau \ll 1$. Now, we have to decompose the $V$ tensor into the new tensors $A'$, $B'$, $C'$ and $R'^{\bigtriangledown}$. This step is called a Higher-Order Singular Value Decomposition (HOSVD). We give the example for the $A'$ tensor. We diagonalize the $dD\times dD$ matrix $W=U \Sigma U^\dagger$~:
\[
W_{(S_A,i);(S_A',i')}=\sum_{S_B,S_C,j,k}\left( V^{S_A',S_B,S_C}_{i',j,k}\right)^* \cdot V^{S_A,S_B,S_C}_{i,j,k}
\]
We keep the $D$ largest eigenvalues of the decomposition in $\tilde{\Sigma}^A$, and $\tilde{U}^A$ is the corresponding $dD\times D$ matrix. The new tensors are $A'^S_{i,j}=\tilde{U}^A_{(S,i);j}/\lambda^{\bigtriangleup A}_{i}$ and $\lambda^{\bigtriangledown A}=\sqrt{\tilde{\Sigma}^A}$ (where $\Sigma ^A \ge 0$ because $W^\dagger=W$). After the updates on the three sites, we can compute the new simplex tensor as \[R'^{\bigtriangledown}_{i,j,k}=\sum_{\{l,m,n\}}\sum_{\{S_A,S_B,S_C\}}T^{S_A,S_B,S_C}_{l,m,n}\left(\tilde{U}^A_{(S_A,l);i}\right)^*\left(\tilde{U}^B_{(S_B,m);j}\right)^*\left(\tilde{U}^C_{(S_C,n);k}\right)^*\]
\\
The renormalization on up triangles is identical, by doing the change $\bigtriangleup \Longleftrightarrow \bigtriangledown$, and taking $\left(A^{S_A}\right)^T$, $\left(B^{S_B}\right)^T$ and $\left(C^{S_C}\right)^T$.

\section{Expectation values}

\subsection{Simple and Full calculations of the environment}

\begin{figure}
\includegraphics[trim=5mm 145mm 10mm 0mm, clip, scale=0.4]{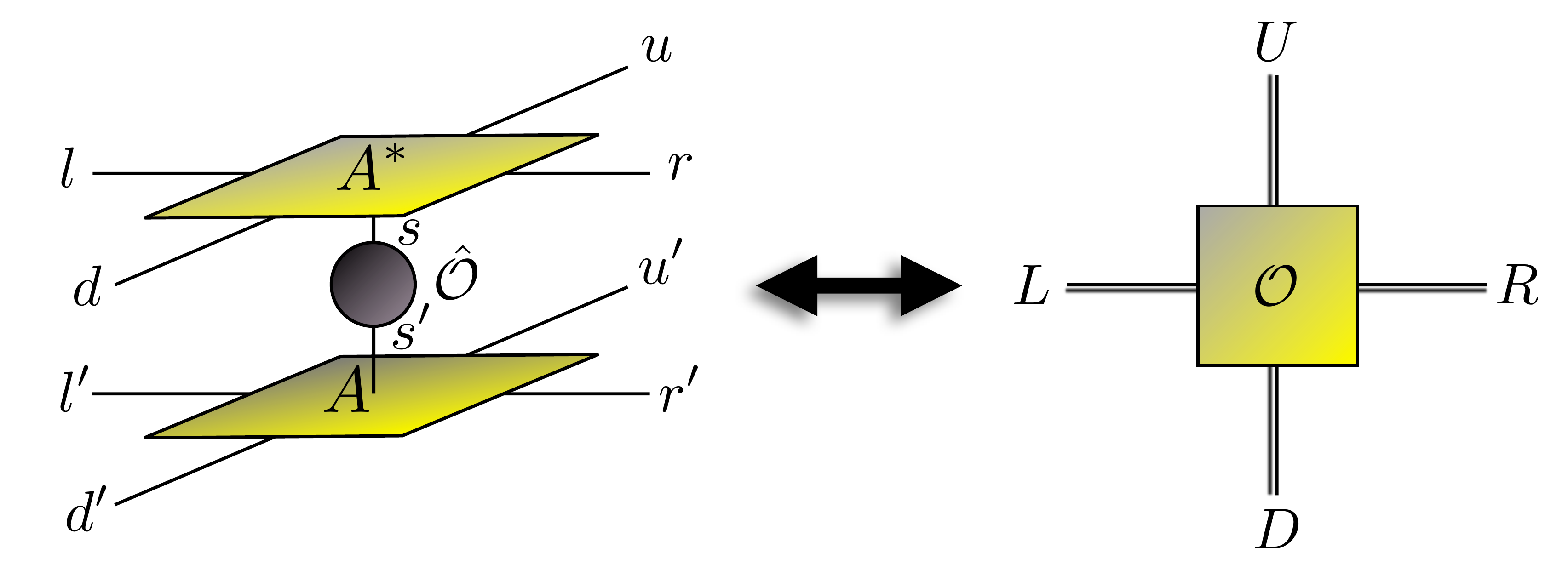}
\caption{Reduced tensor $\mathcal{O}$ of an on-site operator $\hat{\mathcal{O}}$ on an homegeneous square lattice.}
\label{fig:reduced_tensor}
\end{figure}

\begin{figure}
\includegraphics[scale=0.3]{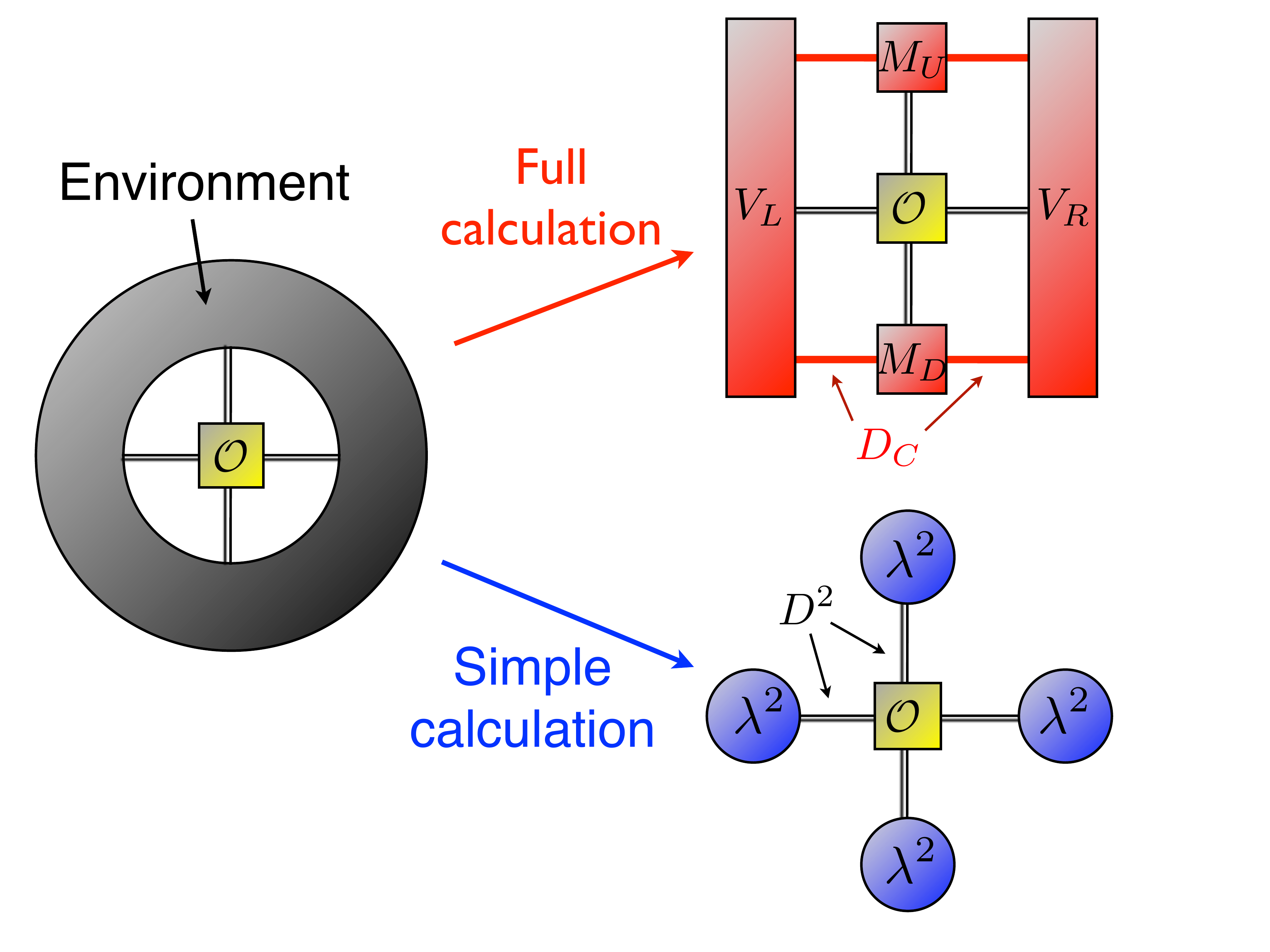}
\caption{The two options to contract the environment of a tensor.}
\label{fig:calculation}
\end{figure}

Usually, to calculate expectation values, one needs to calculate the surrounding system using MPS-based approach, CTMRG or CGTRG methods (see Section above), which requires to reach convergence with the bond dimension $D_C$ of the environment. One simpler approach amounts to use the bond matrices 
$\lambda^{\bigtriangleup(\bigtriangledown)}$ (calculated during the Simple Update optimization) in the same spirit as in one-dimensional system. To illustrate the difference between the two schemes -- which we shall quantitatively compare hereafter -- we consider the example of an on-site operator $\hat{\mathcal{O}}$ on a homogeneous square lattice with a tensor $A$ on each site. First, we compute the reduced tensor $\mathcal{O}_{L,R,U,D}=\sum_{s,s'}(A^s_{l,r,u,d})^*\hat{\mathcal{O}}_{s,s'}A^{s'}_{l',r',u',d'}$, as shown in Fig.~\ref{fig:reduced_tensor}, where $L=(l,l')$, $R=(r,r')$, $U=(u,u')$ and $D=(d,d')$.
After this step, we have to contract the remaining internal bond degrees of freedom with an environment, in order to have a scalar value of $\langle \hat{\mathcal{O}} \rangle$. Two options are then possible and are sketched in Fig.~\ref{fig:calculation}~:

i) On one hand, the Full calculation requires the computation of further tensors. In this example, the MPS-based approach is used and the environment is described by 2 tensors, $M_U$ and $M_D$, of dimension $D_C\times D_C\times D^2$, and 2 vectors, $V_L$ and $V_R$, of dimension $D_C^2D^2$. The new parameter $D_C$ is the bond dimension of the environment and any expectation value, at fixed $D$, must be converged with respect to this parameter $D_c$, as it is not a variational calculation. 

ii) On the other hand, the Simple calculation only requires the computation of $D^2$-dimensional sparse vectors which are $\lambda^2_{(i,j)}=\sum_k\lambda_{k,i}\lambda_{k,j}$, where $\lambda$ is the matrix computed in the Simple update.
In the case of the kagome lattice, using the $T$ tensor of the algorithm, the expectation value is given by:
\[\langle\hat{\mathcal{O}}\rangle=\sum_{\vec{S}',\vec{S}} \sum_{i,j,k} \mathcal{O}_{\vec{S}',\vec{S}} \left(T^{\vec{S}'}_{i,j,k}\right)^* T^{\vec{S}}_{i,j,k}\]
Note that such a scheme becomes in fact exact on the Bethe lattice.
As will be shown next, this Simple calculation of the environment provides very accurate values, in excellent agreement with the Full calculation. Besides, it is much less involved because there is no convergence issue. 

\subsection{Convergence issues in the Full calculation of the environment}

\begin{figure}
\includegraphics[width=\textwidth]{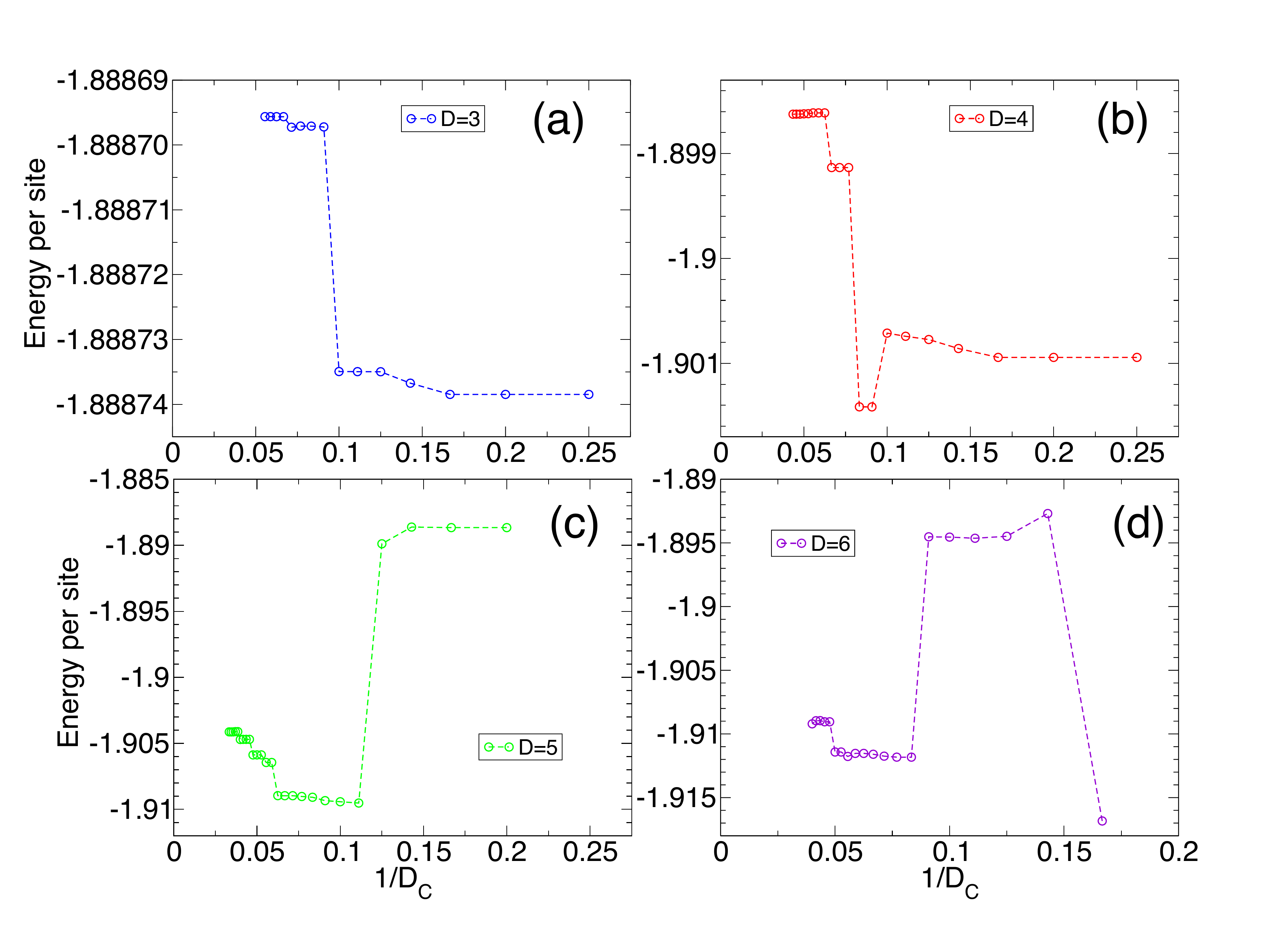}
\caption{Energy per site, using the Full calculation, for different values of $D$, versus the inverse of the environment dimension $D_C$, in the nematic (incompressible) phase ($h=2.6$).}
\label{fig:energies_N}
\end{figure}

\begin{figure}
\includegraphics[width=\textwidth]{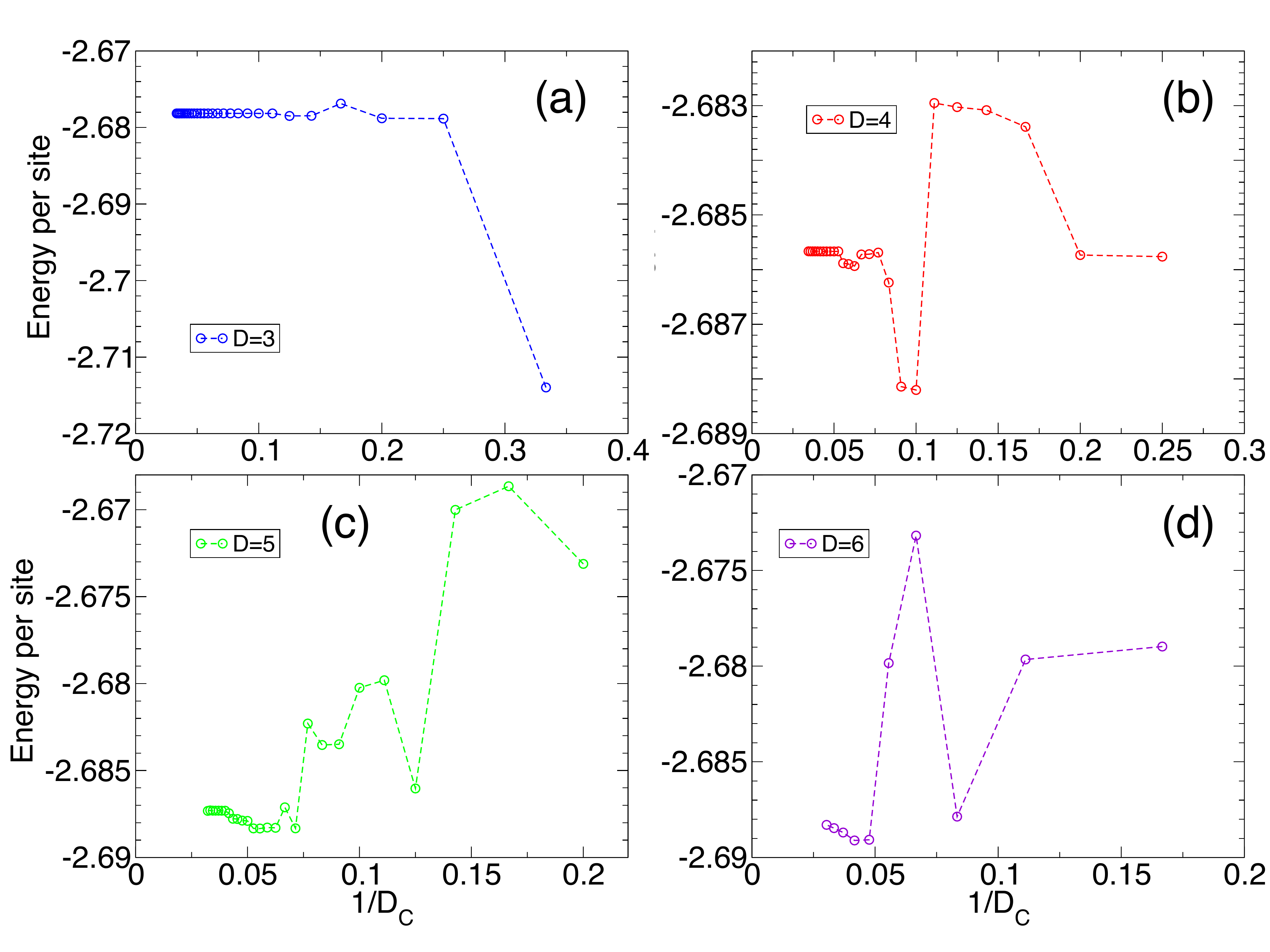}
\caption{Energy per site, using the Full calculation, for different values of $D$ versus the inverse of the environment dimension $D_C$, in a supernematic (compressible) phase ($h=4.29$).}
\label{fig:energies_SN}
\end{figure}

\begin{figure}
\includegraphics[trim=0mm 60mm 0mm 60mm, clip, width=\textwidth]{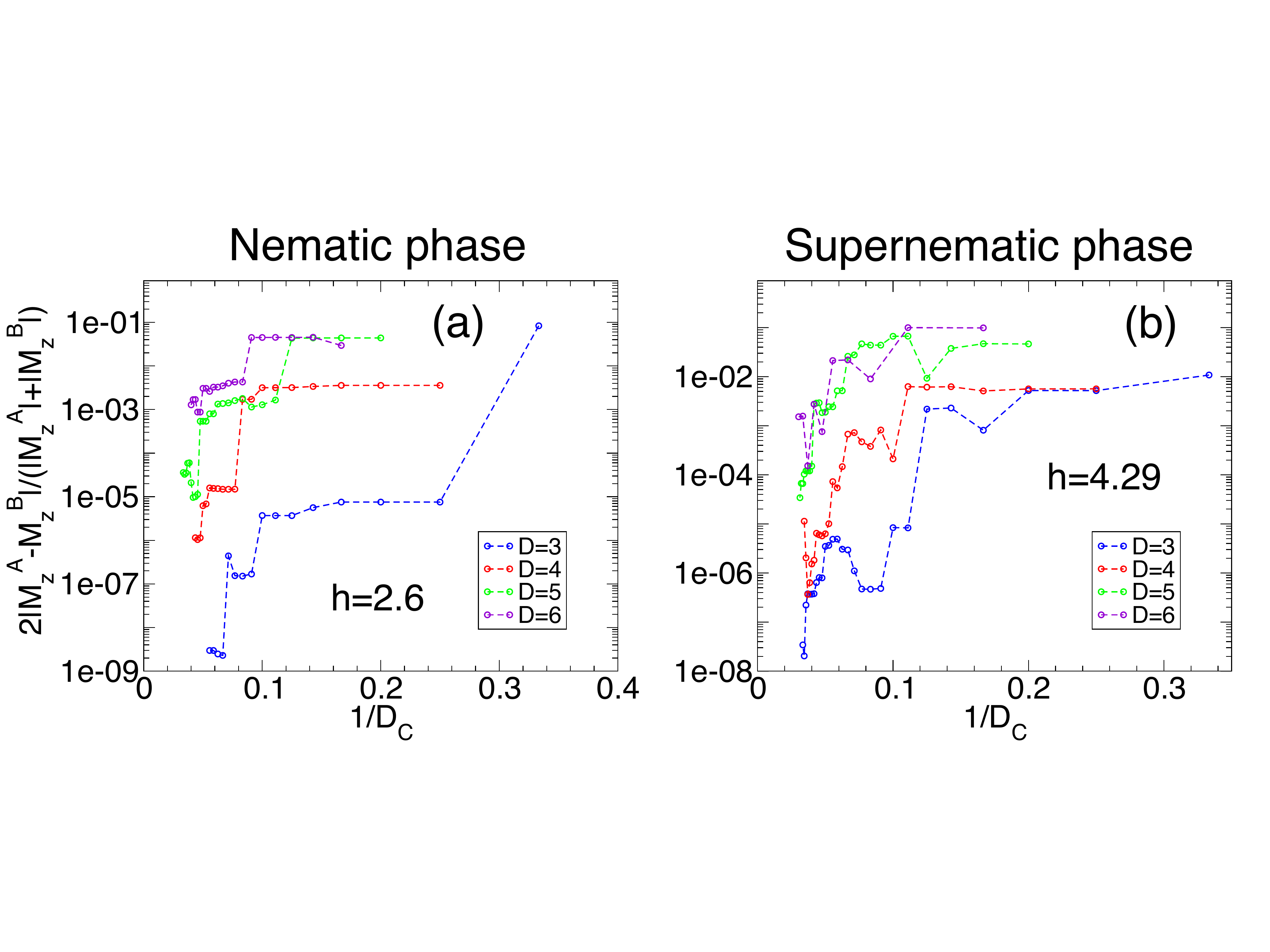}
\caption{Relative difference between $M_z^A$ and $M_z^B$ versus the inverse of the environment bond dimension $D_C$, for the spin-1 Kagome antiferromagnet in the incompressible (a) and compressible phase (b).}
\label{fig:delta}
\end{figure}

In the following, we select two points in the Spin-1 Heisenberg phase diagram, one in the incompressible nematic phase ($h=2.6$) and the other in a compressible supernematic phase ($h=4.29$). 
We first carefully investigate the convergence of various observable with $D_C$ in the Full calculation of the environment. 
In Fig~\ref{fig:energies_N} and Fig~\ref{fig:energies_SN}, we calculate the energy per site $e_0$ with the MPS-based approach for different value of $D$, for $h=2.6$ and $h=4.29$, respectively. We can see the convergence of $e_0$, for $D=$3, 4, 5 and 6, as a function of the inverse of $D_C$. First of all, we notice that it is challenging to perform an extrapolation due to the irregular behavior of the energy when $D_C$ increases. However, we estimate that the energy is converged when one reaches a ``plateau", and the ``jump" between two ``plateaux" is small enough. We can also check the relative difference of the magnetization along the magnetic field between sites A and B, namely $\Delta=2|M_z^A-M_z^B|/(|M_z^A|+|M_z^B|)$, which is 
$\Delta=0$ in the Simple calculation. In Fig~\ref{fig:delta}, this relative difference is plotted versus $1/D_C$, and its value indeed seems to converge to zero when $1/D_C \rightarrow 0$. Interestingly, the convergences of $e_0$ and $\Delta$ with increasing $D_C$ seem to be correlated, with $\Delta < 10^{-4}$ for a converged energy.

\subsection{Comparison of expectation values with Simple or Full calculation of the environment}

\begin{figure}

\includegraphics[trim=0mm 60mm 0mm 60mm, clip, width=\textwidth]{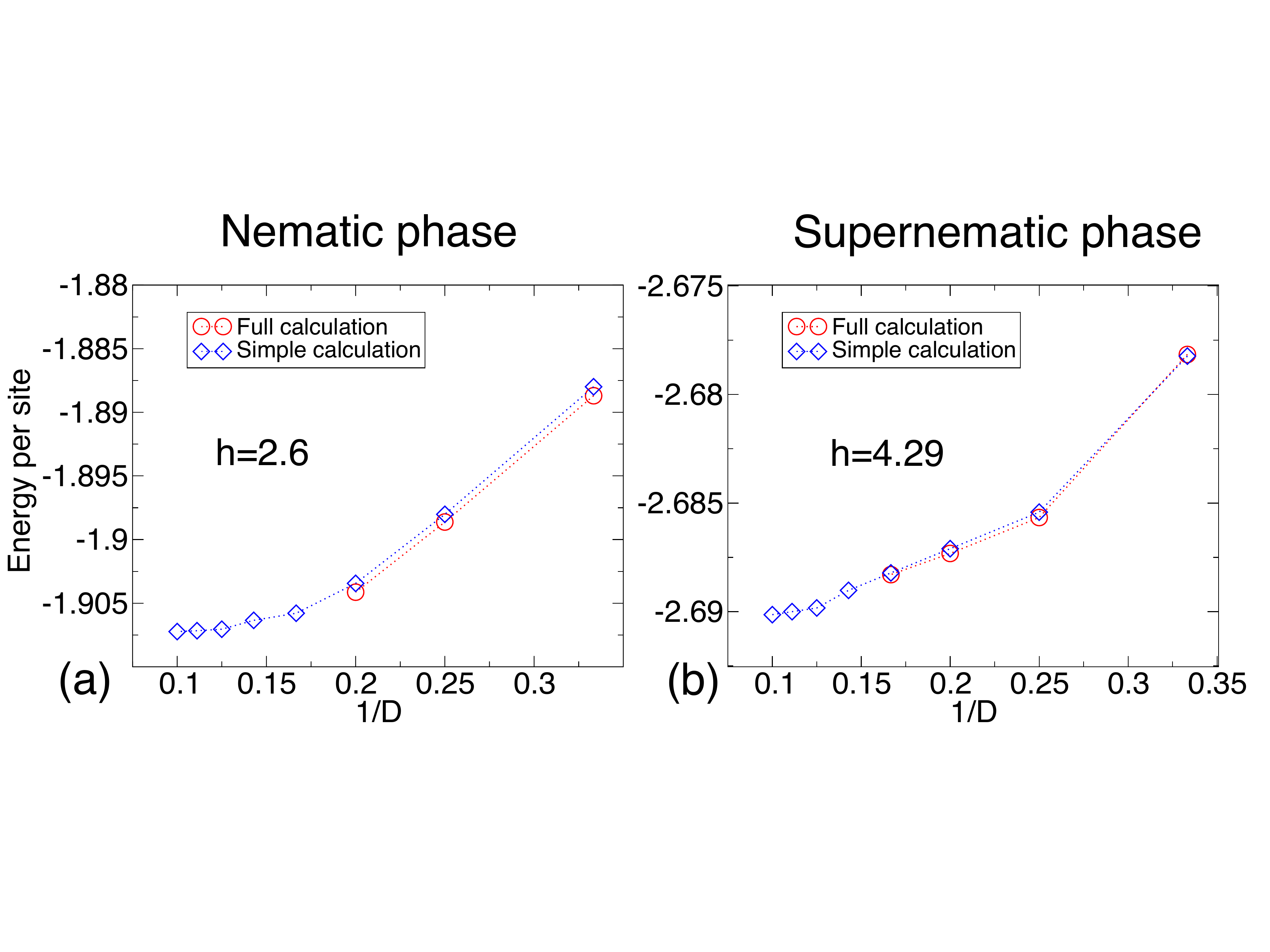}
\caption{Energy per site $e_0$ versus bond dimension $D$, in the simple case (as presented in this Letter) and with the environment (Full) calculation, for the spin-1 Kagome antiferromagnet in the incompressible (a) and compressible phase (b).}
\label{fig:energies}
\end{figure}

Finally, we compare the two methods for both phases (see Fig~\ref{fig:energies}). They are in good agreement, and the relative difference of the energy per site for each selected $D$ is $\delta\simeq 10^{-4}$. 
